\PassOptionsToPackage{bookmarks=false}{hyperref}
\documentclass[sigconf]{acmart}
\pdfoutput=1
\usepackage{booktabs}
\usepackage{amssymb}
\usepackage{amsmath}
\usepackage{xfrac}
\usepackage{gensymb}
\usepackage{url}
\usepackage[utf8]{inputenc}

\usepackage[binary-units=true]{siunitx}
\DeclareSIUnit\transaction{T}

\begin{document}

\copyrightyear{2018} 
\acmYear{2018} 
\setcopyright{acmlicensed}
\acmConference[ICCAD '18]{IEEE/ACM INTERNATIONAL CONFERENCE ON COMPUTER-AIDED DESIGN}{November 5--8, 2018}{San Diego, CA, USA}
\acmBooktitle{IEEE/ACM INTERNATIONAL CONFERENCE ON COMPUTER-AIDED DESIGN (ICCAD '18), November 5--8, 2018, San Diego, CA, USA}
\acmPrice{15.00}
\acmDOI{10.1145/3240765.3240808}
\acmISBN{978-1-4503-5950-4/18/11}

\title{Fast FPGA Emulation of Analog Dynamics in Digitally-Driven Systems}

\author{Steven Herbst}
\affiliation{%
  \institution{Stanford University}
}
\email{sherbst@stanford.edu}
\author{Byong Chan Lim}
\affiliation{%
  \institution{Stanford University}
}
\email{bclim@stanford.edu}
\author{Mark Horowitz}
\affiliation{%
  \institution{Stanford University}
}
\email{horowitz@stanford.edu}

\begin{abstract}
In this paper, we propose an architecture for FPGA emulation of mixed-signal systems that achieves high accuracy at a high throughput.  We represent the analog output of a block as a superposition of step responses to changes in its analog input, and the output is evaluated only when needed by the digital subsystem.  Our architecture is therefore intended for digitally-driven systems; that is, those in which the inputs of analog dynamical blocks change only on digital clock edges.  We implemented a high-speed link transceiver design using the proposed architecture on a Xilinx FPGA.  This design demonstrates how our approach breaks the link between simulation rate and time resolution that is characteristic of prior approaches.  The emulator is flexible, allowing for the real-time adjustment of analog dynamics, clock jitter, and various design parameters.  We demonstrate that our architecture achieves 1\% accuracy while running 3 orders of magnitude faster than a comparable high-performance CPU simulation.
\end{abstract}

\ccsdesc[500]{Hardware~Simulation and emulation}

\keywords{FPGA emulation, mixed-signal circuits, verification}

\maketitle

\section{Introduction}
Top-level simulation is a crucial part of the verification of today's complex chips. For entirely digital designs, FPGA emulation can provide a significant performance boost; gains of 100,000x as compared to CPU simulation have been reported \cite{Asaad:2012}.  However, for systems containing mixed-signal components, as most SoCs do today, emulating analog behavior poses a special challenge: not only does one need to create functional models for analog blocks, but those models must be written in a way that can be implemented on an FPGA.

While there have been many approaches for functional modeling of analog blocks in a digital validation environment, for example using s-domain models~\cite{XMODEL,Jang:2013}, piecewise-linear waveforms~\cite{Lim:2016}, and mixed-mode simulation~\cite{Kundert:2013}, these methods do not map easily onto an FPGA.  Instead, prior work in mixed-signal emulation has represented analog blocks using oversampled discrete-time models~\cite{Nothaft:2014,Deepaksubramanyan:2005,Tertel:2017,Bruckner:2012}.  These models are implemented as infinite impulse response (IIR) and finite impulse response (FIR) filters,  and once their values are quantized, the resulting discrete-time, discrete-value digital filters can be directly mapped onto an FPGA.  While this approach enables emulation of analog circuits, it unfortunately links simulation accuracy to the time step used for the analog blocks.  For systems that use high-speed links, fine time resolution is required to model jitter, meaning that the emulator time step must be much shorter than the shortest clock period in the system, wasting resources and slowing down the system emulation.

To avoid these issues, this paper demonstrates an alternate approach that does not rely on oversampled models, providing accurate emulation results while only using the existing clocks of the system. 

Our approach leverages the fact that most analog blocks in digital systems have inputs that were originally created by another digital block (e.g., a link transmitter, DAC, etc.) before being processed by analog circuits. Thus, in addition to using an event-driven, variable time step approach like~\cite{Lim:2016}, we further accelerate emulation by eliminating the need to create internal analog events. Instead, emulated time progresses directly from one emulated clock edge to the next.  Analog outputs are computed as a superposition of step responses to changes in analog inputs that are digitally-driven, meaning that they change only on digital clock edges.  We demonstrate that the accuracy of our approach is independent of the time step.

The proposed architecture is presented in Section~\ref{sec:arch}, followed by an implementation example in Section~\ref{sec:impl} for an \SI{8}{\giga\transaction/\second} high-speed link transceiver.  Measured performance of our architecture on a Xilinx FPGA is reported in Section~\ref{sec:results}, and Section~\ref{sec:extension} covers possible extensions to handle nonlinearity and a broader class of inputs.

\section{Architecture}
\label{sec:arch}

\begin{figure*}
\centering
\includegraphics[width=0.81\textwidth]{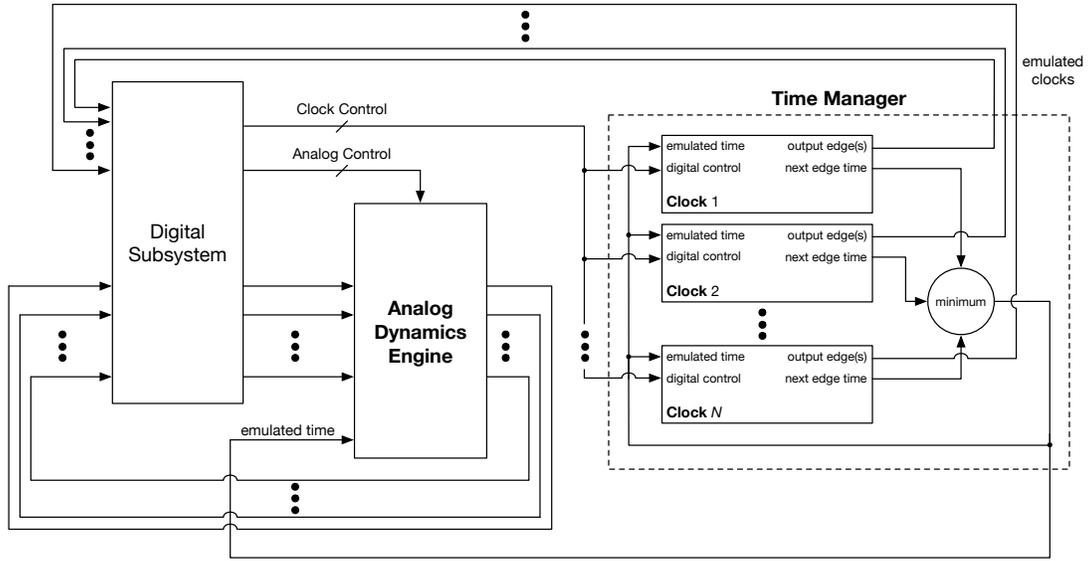}
\caption{The proposed emulator architecture.  A digital subsystem generates piecewise-constant analog inputs for an analog dynamics engine (ADE) whose outputs are computed as superpositions of pulse responses.  A time manager moves emulated time forward from one emulated clock edge to the next and generates clocks for the digital subsystem.}
\label{fig:toplevel}
\end{figure*}

\begin{figure*}
\centering
\includegraphics[width=0.95\textwidth]{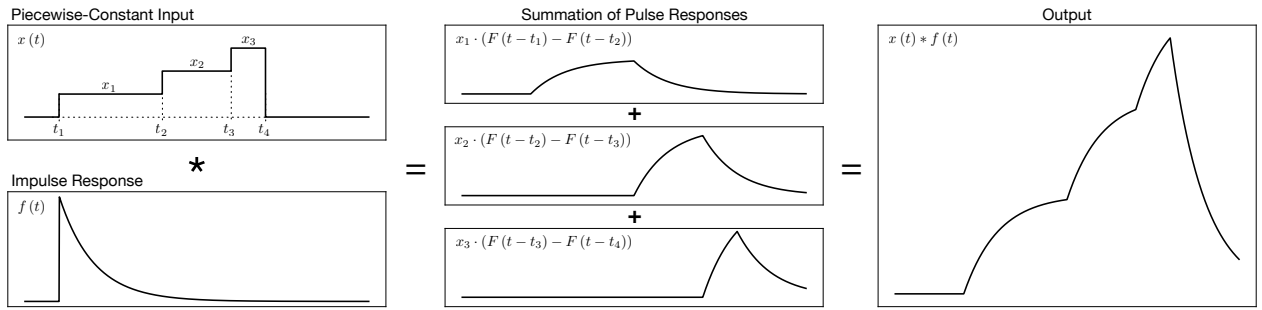}
\caption{The response of a linear time-invariant (LTI) system to a piecewise-constant input can be computed as a summation of pulse responses, each dependent on two values of the system's step response $F\left(t\right)$.}

\label{fig:pwc2out}
\end{figure*}

Figure~\ref{fig:toplevel} shows our emulation architecture, in which a digital subsystem interacts with analog blocks through an analog dynamics engine (ADE) and a time manager.  The ADE transforms digitally-driven analog inputs into analog outputs, while the time manager determines the time associated with each emulator cycle and generates clock edges for the digital subsystem.  The architecture is intended to be implemented entirely by the programmable logic of an FPGA, avoiding the need for an analog daughtercard.

\subsection{Analog Dynamics Engine}

In a conventional mixed-signal emulator, each emulation cycle corresponds to a fixed time step $\Delta t$, with analog dynamics mapped to hardware as FIR and/or IIR filters.  However, the accuracy of such an approach generally worsens as $\Delta t$ increases.  For example, if Euler's method is used to generate the filter coefficients, the global truncation error (GTE) is $\mathcal{O}\left(\Delta t\right)$, while the GTE associated with the trapezoid rule is $\mathcal{O}\left(\Delta t^2\right)$~\cite{Suli:2003}.  As a result, there is a trade-off between emulation rate and accuracy: high accuracy can be achieved, but at a low throughput, and vice versa.

In our architecture, every emulation cycle corresponds to one or more digital clock edges.  Under the assumption that analog blocks are digitally-driven, these are the exact times at which analog inputs might change.  The ADE therefore has a complete history of the precise times and values of analog input steps, enabling each analog output to be computed as a summation of step responses.

Our approach assumes that the analog blocks being modeled are linear and time-invariant (LTI).  Section~\ref{sec:impl} describes how this approach can be simply extended to handle time-variant systems, and Section~\ref{sec:extension} discusses how nonlinearity could be handled.

It is well known that an LTI system is characterized by an impulse response $f\left(t\right)$, with its output $y\left(t\right)$ equal to the convolution of its input $x\left(t\right)$ with $f\left(t\right)$:
\begin{equation}
\label{eqn:ydef}
y\left(t\right) = f\left(t\right) \ast x\left(t\right) = \int_{-\infty}^{\infty} f\left(\tau\right) \cdot x\left(t-\tau\right)\,d\tau 
\end{equation}

When the input changes only on digital clock edges, the integral can be reduced to a summation.  Suppose that $x\left(t\right)$ is a general piecewise-constant function:
\begin{equation}
\label{eqn:xdef}
x\left(t\right) = \sum_k x_k \cdot \left(u\left(t-t_k\right)-u\left(t-t_{k+1}\right)\right)
\end{equation}
where $u\left(t\right)$ is the unit step function.  Substituting Equation~\ref{eqn:xdef} into Equation~\ref{eqn:ydef} yields:
\begin{equation}
\label{eqn:sumpulses}
\begin{split}
y\left(t\right) &= \sum_k x_k\cdot\int_{t-t_{k+1}}^{t-t_{k}} f\left(\tau\right)\,d\tau \\
&= \sum_k x_k \cdot \left(F\left(t-t_{k}\right)-F\left(t-t_{k+1}\right)\right)
\end{split}
\end{equation}
where $F\left(t\right)$ is the system's step response:
\begin{equation}
F\left(t\right) = \int_{0}^{t} f\left(\tau\right)\,d\tau 
\end{equation}

Equation~\ref{eqn:sumpulses} can be interpreted as a summation of pulse responses weighted by the sequence of input values $x_k$, as illustrated in Figure~\ref{fig:pwc2out}.  Since each pulse response is simply the difference between two values of $F\left(t\right)$, the step response can be precomputed once and subsequently used to calculate the system output for any piecewise-constant input.

As long as the analog input is digitally-driven and the system is LTI, Equation~\ref{eqn:sumpulses} is exact, meaning that the accuracy of our approach does not depend on the size of the time steps taken by the emulator.  Notice that since our system makes no assumptions about the width of input pulses, the effects of jitter on the clocks driving analog blocks will automatically be included in the analog outputs. In addition, there is no need to approximate analog dynamics by a rational transfer function, as in~\cite{XMODEL}, since our architecture makes direct use of the system's step response.  This is particularly convenient when working with measured frequency response data such as the S-parameter model of a backplane channel \cite{Palermo:2017}.

\subsection{Time Manager}

The time manager has two tasks: 1) it assigns a time to each emulation cycle, and 2) it generates clock edges for emulated blocks.  To maximize the emulation throughput, its objective is to take the largest time step possible without skipping over any clock edges.

In our architecture, every emulated clock stores the time of its next clock edge, and the minimum of these times is selected for the next emulation cycle.  Any clock whose edge is to occur at that time will output the edge and update the time of its next edge.

Our time manager design ensures that at least one clock edge is generated in every emulation cycle; there are no ``analog-only'' cycles.  As a result, the emulation rate of the entire mixed-signal system is similar to that of the digital subsystem alone.

\begin{figure*}
\centering
\includegraphics[width=0.91\textwidth]{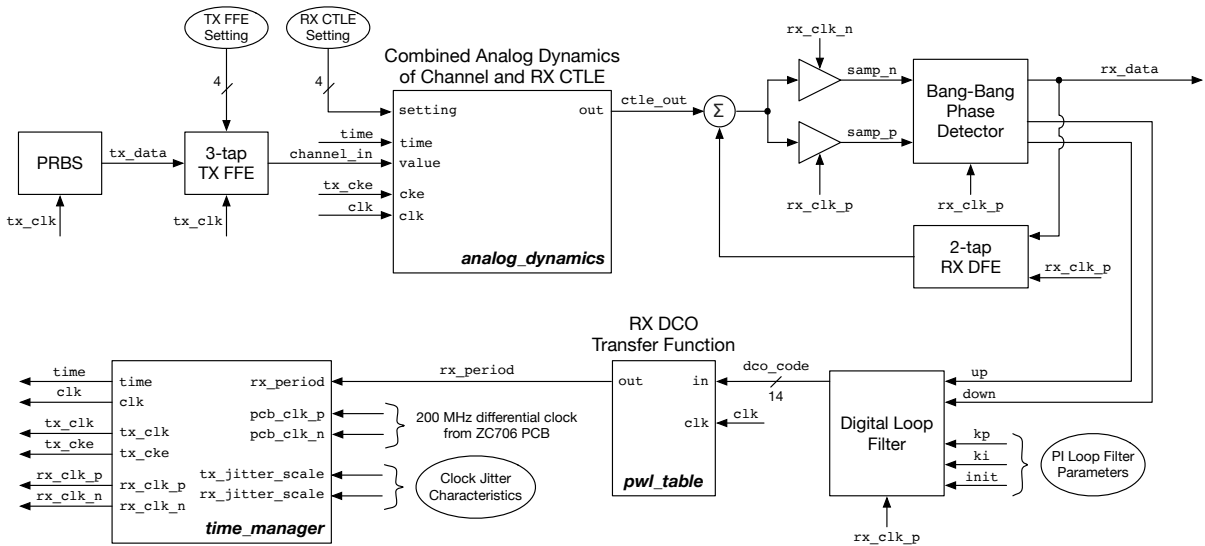}
\caption{Block diagram of our FPGA emulator for a high-speed link transceiver.  The digitally-driven \texttt{channel\char`_in} signal is the input to the analog dynamics engine (ADE), which implements the combined dynamics of the channel and an adjustable continuous-time linear equalizer (CTLE).  The \texttt{time\char`_manager} module controls the flow of time in the emulator and determines which clocks are active in each emulation cycle.  The nonlinear transfer function of the digitally-controlled oscillator (DCO) is implemented by a piecewise-linear lookup table.  Circled settings are adjustable in real-time by users.}
\label{fig:sysdiagram}
\end{figure*}

\begin{figure}
\centering
\includegraphics[width=0.85\columnwidth]{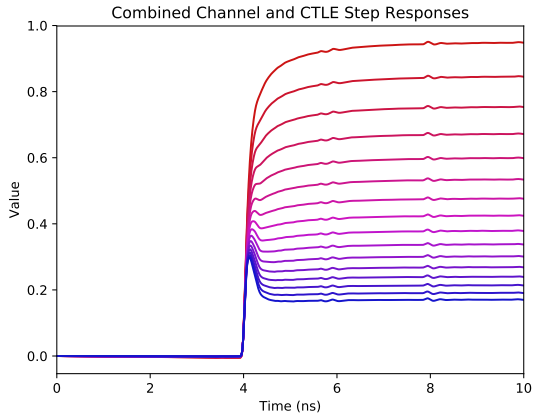}
\caption{Family of step responses representing analog dynamics in the high-speed link model.  Each step is the convolution of a channel step response, computed from measured S-parameters, with the impulse response of the RX CTLE in one of 16 different settings.  In our FPGA emulator, the CTLE dynamics setting can be changed in real-time, without reprogramming the FPGA.}
\label{fig:stepfam}
\end{figure}

\section{Implementation}
\label{sec:impl}

We implemented our emulator architecture for an \SI{8}{\giga\transaction/\second} high-speed link transceiver design.  As shown in Figure~\ref{fig:sysdiagram}, our emulator has adjustable transmitter (TX) and receiver (RX) equalization, with a clock and data recovery (CDR) loop that is closed through a bang-bang phase detector (BBPD) and digitally-controlled oscillator (DCO).  A decision-feedback equalizer (DFE) helps to reduce intersymbol interference (ISI).

\subsection{Precomputation of Analog Dynamics}

The ADE implements the combined analog dynamics of the channel and the continuous-time linear equalizer (CTLE).  Channel dynamics were computed from S-parameter measurements \cite{Palermo:2017}, while the CTLE transfer function was based on the PCIE specification, consisting of two fixed poles and an adjustable zero to support adaptive equalization.

In our implementation, the CTLE zero can be positioned in one of 16 different settings between \SIrange{0.4}{2.0}{\giga\hertz}, meaning that the CTLE is not time-invariant.  To handle this, the ADE stores a family of precomputed step responses, each representing the combined dynamics of the channel and CTLE in one of its settings (Figure~\ref{fig:stepfam}).  During each emulation cycle, the ADE selects the appropriate step response based on the current CTLE setting.

The dynamics of transitions between CTLE settings is not captured by this approach.  However, it is often unnecessary to model these transitions in detail, since they are typically much shorter than the amount of time that the adaptive equalization algorithm spends in a given CTLE setting.

\begin{figure*}
\centering
\includegraphics[width=0.88\textwidth]{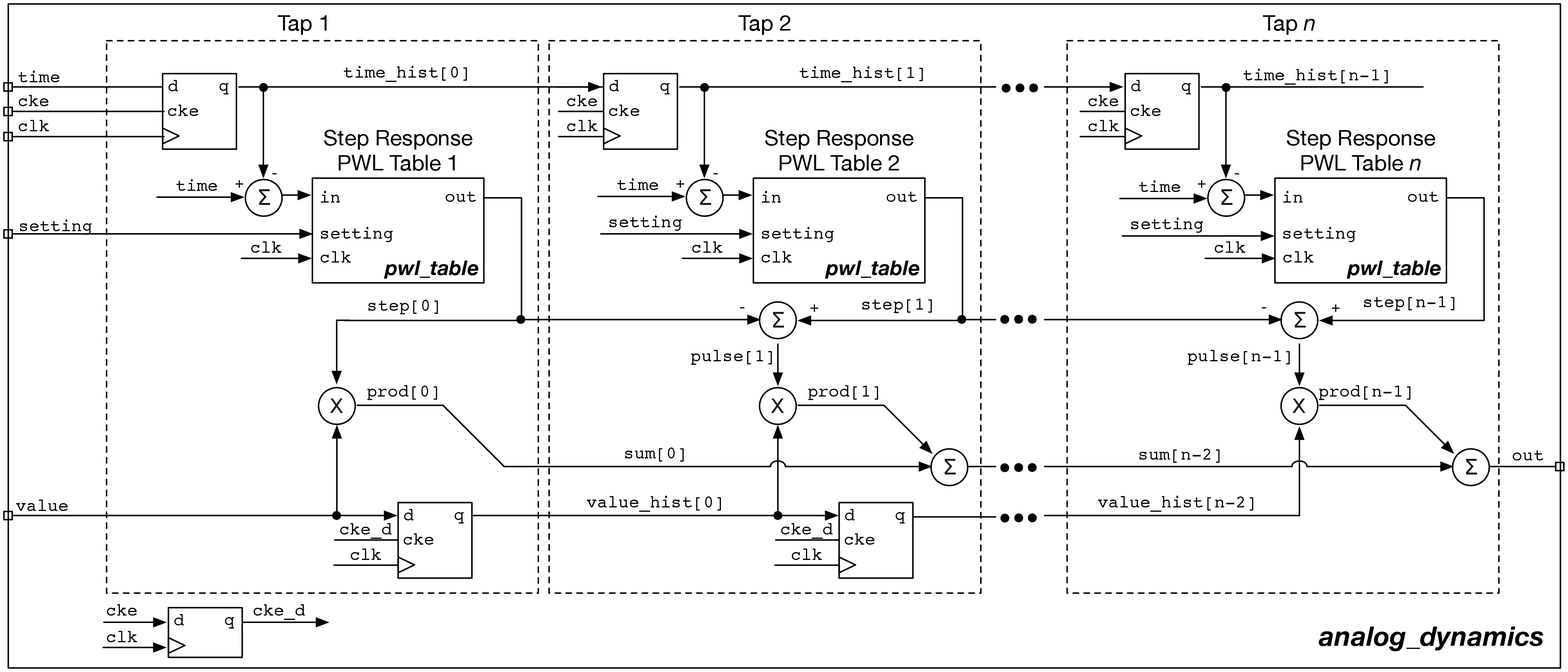}
\caption{Implementation of the analog dynamics engine (ADE).  Each pulse response is computed as the difference between two step response values, which depend on the timing of step changes in the analog input.  The pulse responses are weighted by values from the input history and summed to form the ADE output.  Our emulator uses 85 taps to store an input history spanning about 10\;ns, with each tap corresponding to a unit interval (UI) of the link.}
\label{fig:anadyn}
\end{figure*}

\subsection{Analog Dynamics Engine}

As shown in Figure~\ref{fig:anadyn}, the ADE is implemented as an array of taps, each containing a step response lookup table.  Following Equation~\ref{eqn:sumpulses}, the ADE output is a weighted sum of its input history, with each weight computed as the difference in step response value between two neighboring ADE taps.

Since the ADE has a finite number of taps, the implementation effectively truncates Equation~\ref{eqn:sumpulses}.  For our system, 85 taps was sufficient to limit the truncation error to a few tenths of a percent.  As a guideline, the number of taps should be approximately the ratio of the step response settling time to the TX clock period, independent of the emulation time resolution.

\subsection{Compressing Step Response Data}
\label{sec:pwlcompr}

To reduce the memory footprint of the ADE, each of its taps is compressed by trimming its domain and by using piecewise-linear approximations.

\subsubsection{Domain Trimming}

In general, each ADE tap must evaluate the step response over a different timespan.  For example, the first tap in our ADE reads out step response values near ${t=\SI{0}{\nano\second}}$ while the last tap reads out values near ${t=\SI{10}{\nano\second}}$.  More generally, the $k$th ADE tap will evaluate the step response at a time between ${\left(k\!-\!1\right)}$ and $k$ periods of the TX clock.  

Taking advantage of this property, we trim the step response lookup tables individually so that taps store only the data they may actually need.  In our emulator, the TX clock has a nominal period $T_{\scriptscriptstyle TX}$ and a period jitter that is uniformly distributed between $-J_{\scriptscriptstyle TX}$ and $+J_{\scriptscriptstyle TX}$, so its period is guaranteed to lie between $T_{\scriptscriptstyle TX}-J_{\scriptscriptstyle TX}$ and $T_{\scriptscriptstyle TX}+J_{\scriptscriptstyle TX}$.  Hence, the domain of the $k$th lookup table can be trimmed to the following interval:
\begin{equation}
\left(k\!-\!1\right)\cdot\left(T_{\scriptscriptstyle TX}\!-\!J_{\scriptscriptstyle TX}\right) \leq t - t_k \leq k\cdot\left(T_{\scriptscriptstyle TX}\!+\!J_{\scriptscriptstyle TX}\right)
\end{equation}

If the jitter distribution were unbounded, domain trimming would require an approximate time range to be determined for each tap.  For example, suppose that the TX clock periods were modeled as independent and identically distributed Gaussian random variables with mean $T_{\scriptscriptstyle TX}$ and standard deviation $\sigma_{\scriptscriptstyle TX}$.  The sum of $n$ periods would then be a Gaussian random variable with mean $nT_{\scriptscriptstyle TX}$ and standard deviation $\sigma_{\scriptscriptstyle TX}\sqrt{n}$.

Hence, with high probability ($1\!-\!2\!\cdot\!10^{-9}$), the duration of $n$ TX periods would be $nT_{\scriptscriptstyle TX}\pm 6\sigma_{\scriptscriptstyle TX}\sqrt{n}$.  It would therefore likely be acceptable to trim the domain of the $k$th lookup table to the following interval:
\begin{equation}
\left(k\!-\!1\right)T_{\scriptscriptstyle TX}\!-\! 6\sigma_{\scriptscriptstyle TX}\sqrt{k\!-\!1} \leq t - t_k \leq kT_{\scriptscriptstyle TX}\!+\! 6\sigma_{\scriptscriptstyle TX}\sqrt{k}
\end{equation}
In the exceptional case that an ADE tap must evaluate the step response outside of this range, it could do so by extrapolation or by reading from a neighboring tap.

\subsubsection{Piecewise-Linear Approximation}

\begin{figure}
\centering
\includegraphics[width=0.89\columnwidth]{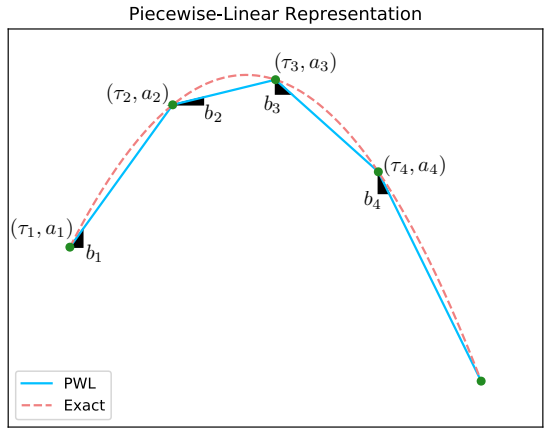}
\caption{Our emulator uses piecewise-linear lookup tables to store step responses and nonlinear functions in a compressed form.  Each time point $\tau_k$ in the lookup table is assigned an offset $a_k$ and a slope $b_k$.}
\label{fig:pwldef}
\end{figure}

We use piecewise-linear (PWL) lookup tables to store step response data in a memory-efficient manner.  As illustrated in Figure~\ref{fig:pwldef}, step responses were approximated by a sequence of line segments whose offsets and slopes were stored in lookup tables.  For each ADE tap, the number of PWL segments was determined by starting with two segments and iteratively doubling the number of segments until the error in the PWL representation was less than \SI{0.1}{\percent}.  Within each iteration, linear programming was used to determine an optimal PWL representation.

While not done in this current implementation, a multivariate PWL representation could be used to further reduce the memory overhead of an adjustable model by interpolating between its settings instead of storing the step response of each one.

\subsection{Time Manager}

There are two emulated clocks in our system, one for the transmitter and one for the receiver.  The RX clock has two output phases, since the BBPD uses both edges of the RX clock, while the TX clock has only one output phase to represent its rising edge.

\begin{figure}
\centering
\includegraphics[width=0.88\columnwidth]{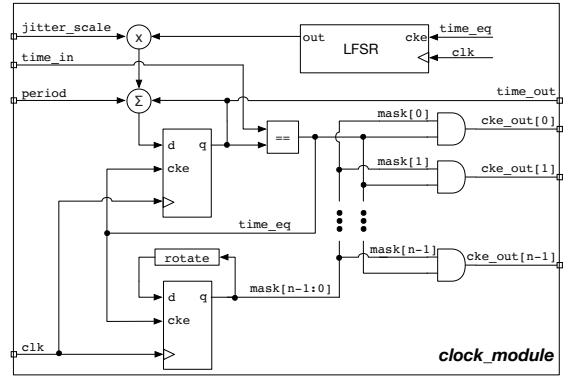}
\caption{Clock module used to model TX and RX clock behavior.  During each emulation cycle, \texttt{time\char`_in} is compared to \texttt{time\char`_out}.  If they match, \texttt{time\char`_out} is incremented and one of the \texttt{cke\char`_out} signals is asserted.  The clock phase is advanced by rotating \texttt{mask}, and jitter is implemented by adding the scaled output of an LFSR to the \texttt{period} input.}
\label{fig:clkimpl}
\end{figure}

Both clocks are implemented using instances of the module shown in Figure~\ref{fig:clkimpl}.  During each emulation cycle, the clock module compares the emulation time, \texttt{time\char`_in}, to the time of its next clock edge, \texttt{time\char`_out}.  If the two match, it asserts one of the \texttt{cke\char`_out} signals and increments \texttt{time\char`_out}.  The increment includes jitter, which is implemented by scaling the output of a linear feedback shift register (LFSR).

The \texttt{cke\char`_out} signals are used to generate clock signals for each clock phase by gating a free-running \SI{30}{\mega\hertz} emulator clock on a cycle-by-cycle basis.  The gating itself is performed by a Xilinx Mixed-Mode Clock Manager (MMCM) IP Block, which ensures glitch-free gating and low inter-clock skew.

\section{Results}
\label{sec:results}

In this section, we discuss measurements of emulator throughput, resource utilization, and accuracy.  The Xilinx ZC706 board, which features a Xilinx Zynq-7045 FPGA, was used in these evaluations.  Compiling our high-speed link emulator for this platform took \SI{11}{\minute} from synthesis to bitstream generation using Vivado 2016.4.

\subsection{Emulation Throughput}

Table~\ref{tab:emperf} summarizes the performance of our high-speed link emulator in comparison to optimized CPU simulations  of similar systems.  Since the emulator clock rate is \SI{30}{\mega\hertz}, and three emulator clock cycles are required to process each unit interval (UI), the emulator throughput is \SI{e7}{UI/\second}, or \SI{1.25}{\milli\second/\second}.

\subsubsection{Performance Comparison} The CPU simulation rate of a similar high-speed link written in SystemVerilog as a fast functional model for validation was \SIrange{0.741}{1.429}{\micro\second/\second}~\cite{Lim:2016}.  In this model, an event-driven approach was used, with analog dynamics represented as a sum of linear filters and waveforms represented by piecewise-linear segments. Our emulation system is at least 875x faster than this comparable, high-performance CPU simulation.

A different fast CPU simulation approach based on s-domain modeling has also been reported~\cite{XMODEL}.  In that case, a high-speed link was implemented using a 50-pole channel model, and a simulation rate of \SIrange{0.110}{0.129}{\micro\second/\second} was achieved.  In comparison, our emulation throughput is at least 9,690x higher.

In order to isolate the speedup in running our emulator on an FPGA, we measured the performance of a CPU-based simulation of the emulator's SystemVerilog code.  This simulation was run using Cadence Xcelium 18.03 on an Intel Xeon E5645 CPU (\SI{2.4}{\giga\hertz}) with \SI{96}{\giga\byte} RAM; we measured a simulation rate of \SI{0.192}{\micro\second/\second}.  Hence, our architecture runs 6,510x faster on an FPGA than on a multi-core CPU-based simulator.

\begin{table}
\centering
\caption{Emulator performance in comparison to optimized simulations of similar systems.}
\label{tab:emperf}
\begin{tabular}{ccc >{\bfseries}c}
\toprule
Design & \cite{XMODEL} & \cite{Lim:2016} & This Work\\
\midrule
Type & CPU & CPU & FPGA \\
Simulation Rate & 0.129\;µs/s & 1.429\;µs/s & 1.250\;ms/s \\
Speedup vs. \cite{XMODEL} & 1x & 11x & 9,690x \\
\bottomrule
\end{tabular}

\bigskip

\caption{Resource utilization of the entire emulator.}
\label{tab:allres}
\begin{tabular}{cccc}
\toprule
Resource &Count & Available & Percent Utilization\\
\midrule
LUT & 11,392 & 218,600 & \SI{5.2}{\percent} \\
FF & 8,076 & 437,200 & \SI{1.8}{\percent} \\
BRAM & 93 & 545 & \SI{17.1}{\percent} \\
DSP & 141 & 900 & \SI{15.7}{\percent} \\
\bottomrule
\end{tabular}

\bigskip

\caption{Resource utilization of the analog dynamics engine (ADE).}
\label{tab:aderes}
\begin{tabular}{cccc}
\toprule
Resource&Count&Available&Percent Utilization\\
\midrule
LUT & 8,085 & 218,600 & \SI{3.7}{\percent} \\ 
FF & 3,475 & 437,200 & \SI{0.8}{\percent} \\ 
BRAM & 36 & 545 & \SI{6.6}{\percent} \\ 
DSP & 138 & 900 & \SI{15.3}{\percent} \\ 
\bottomrule
\end{tabular}

\end{table}

\subsection{Resource Utilization}

\begin{figure}
\centering
\includegraphics[width=\columnwidth]{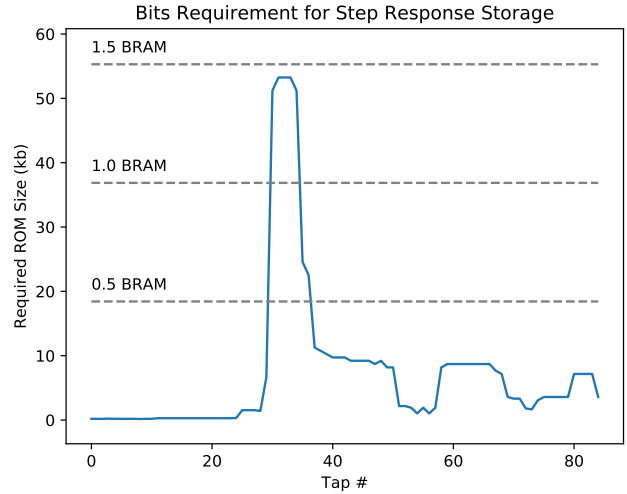}
\caption{Number of bits required to store step response data for each tap in the analog dynamics engine (ADE).  Memory on the Xilinx Zynq-7045 FPGA used in our implementation can be allocated in increments of a half BRAM tile (18\;kb) as indicated by the dashed lines.  Only 8\;\% of taps required more than the minimum increment.}
\label{fig:fpgares}
\end{figure}

Table~\ref{tab:allres} shows the FPGA resource utilization of the entire emulation system, demonstrating that no more than \SI{17.1}{\percent} of any resource was needed.  As a result, there would be ample room left over to emulate a larger digital subsystem. For example, one could build a more complete multi-lane SERDES system including equalization adaptation logic and PCS (Physical Coding Sublayer).

Table \ref{tab:aderes} summarizes the resource utilization of the ADE alone. Note that only \SI{39}{\percent} of block RAM (BRAM) in the emulator is used by ADE;  the remaining BRAM tiles are consumed by Integrated Logic Analyzer (ILA) IP blocks to capture internal waveforms.

\subsubsection{LUTs and FFs}
To put the LUT and FF utilization in perspective, one Xilinx MicroBlaze soft processor consumes 2,071 LUTs and 1,672 FFs with typical settings on our FPGA \cite{MicroBlaze:2018}.  Hence, our emulator has a LUT and FF footprint equivalent to about five or six MicroBlaze cores.  Approximately 100 more such cores could fit in the resources remaining after instantiating our emulator.

\subsubsection{BRAM}

Our individual optimization of the ADE taps reduced the BRAM tile requirement by more than 22.5x, from 2,097 (which would not have fit on our FPGA) to 93.  As a result, the largest consumers of BRAM in our emulator are the ILA blocks, rather than the PWL tables in the ADE.

Figure~\ref{fig:fpgares} shows the optimized lookup table size for each ADE tap.  More memory is required to represent parts of the step response that are rapidly varying, such as the area around $t=\SI{4}{\nano\second}$ in Figure~\ref{fig:stepfam}. However, the majority of tables (\SI{92}{\percent}) require no more than a half BRAM tile, which is the minimum unit that can be allocated on our FPGA.

\subsubsection{DSP}

Most (\SI{97.9}{\percent}) of the DSP utilization is attributed to the ADE.  Each of its 85 taps has two multipliers: one to implement its PWL table, and one to weight its input value by the corresponding pulse response.  Since each DSP slice in our FPGA contains a single multiplier, the expected number of DSP slices consumed by the ADE is therefore around 170.  The actual DSP utilization is \SI{18.3}{\percent} lower, since some low-precision multiplications synthesized to LUTs.

\subsection{Emulation Accuracy}

The accuracy of our emulator is evaluated in two ways. First, the waveforms of our FPGA emulation are directly compared with those of a CPU simulation.  Second, we compare high-level behavioral metrics between the two approaches.

Compared to the CPU simulation, there are several potential sources of error in the emulation which include: 1) quantization error due to a fixed-point representation, 2) PWL approximation error in representing step responses, and 3) truncation error due to the finite number of pulses used.

\subsubsection{Transient Accuracy} 
\begin{figure}
\centering
\includegraphics[width=0.93\columnwidth]{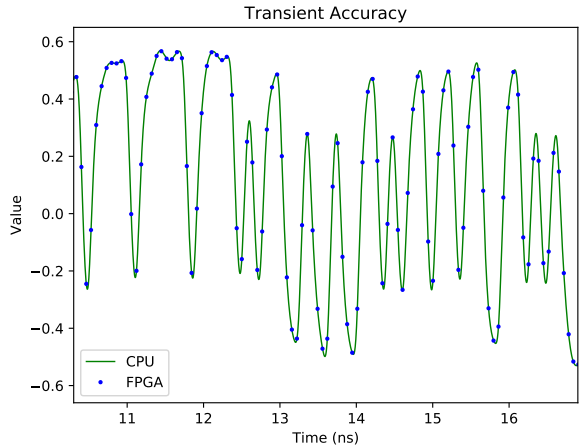}
\caption{Comparison of FPGA emulation and CPU simulation waveforms, shown at the output of the RX CTLE.  Our emulator evaluates analog signals only on digital clock edges, and these samples are shown as dots.  The continuous-time waveform was calculated using CPU simulation and is shown as a solid curve.  Over all 160 combinations of TX and RX equalizer settings, the worst-case relative error observed was \mbox{-0.7\%/+1.1\%.}}
\label{fig:trcheck}
\end{figure}

The ADE output was captured by the ILA and compared with the corresponding output from a CPU simulation.  The waveform comparison was conducted over all 160 different configurations of TX and RX equalization settings.  In each configuration, emulated and simulated waveforms were compared over a \SI{128}{\nano\second} duration (1024 UIs).  The worst-case relative error\footnote{Relative error is defined as $ \mathrm{max} \left| y_{\text{\tiny FPGA}}-y_{\text{\tiny CPU}}\right| / \mathrm{max}\left| y_{\text{\tiny CPU}}\right|$, where $y_{\text{\tiny FPGA}}$ and $y_{\text{\tiny CPU}}$ are the FPGA and CPU waveforms, respectively.} observed was -0.7/+1.1\%.  Figure \ref{fig:trcheck} shows a sample comparison of the transient waveforms.

\subsubsection{Behavioral Accuracy}

\begin{figure}
\centering
\includegraphics[width=0.92\columnwidth]{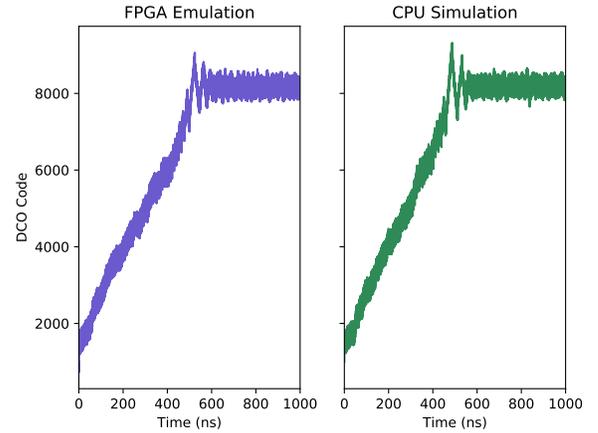}
\caption{Startup transient of the high-speed link transceiver captured by the FPGA emulator.  The RX DCO code is shown over the first 1\;µs of operation, starting from an initial value of 1000 (7.6\;GHz) and settling around 8192 (8.0\;GHz).}
\label{fig:startupcheck}
\end{figure}

\begin{figure}
\centering
\includegraphics[width=0.92\columnwidth]{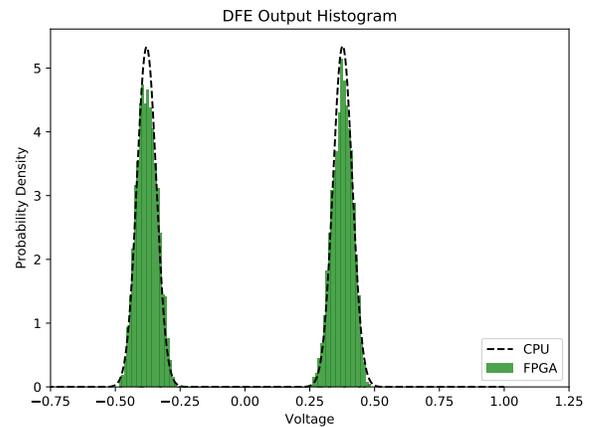}
\caption{Amplitude histogram at the DFE output, constructed from emulator data.  The dashed curve represents the distribution of comparable data gathered in a CPU simulation.}
\label{fig:histcheck}
\end{figure}

Figure~\ref{fig:startupcheck} shows a comparison of the CDR startup waveforms from our emulator and from a CPU simulation.  They are in generally good agreement, with \SI{10}{\percent} settling times of \SI{513}{\nano\second} and \SI{524}{\nano\second} for the emulation and simulation, respectively.  Figure~\ref{fig:histcheck} compares amplitude histograms at the DFE output.  The standard deviation about ``0'' and ``1'' levels was \SI{39.2}{\milli\volt} for the emulation and \SI{37.4}{\milli\volt} for the simulation.  Note that the CPU simulations used in these two comparisons did not include clock jitter.

\section{Extensions}
\label{sec:extension}

In this section, we describe possible extensions to model nonlinearity and handle a broader class of analog input signals.

\subsection{Handling a Broader Class of Inputs}

Up until this point, we have used the term \emph{digitally-driven} to refer to analog input signals that change only on digital clock edges, meaning that they are that are piecewise-constant.  However, it is possible to broaden this definition to include piecewise-polynomial waveforms, which can at least approximately represent any arbitrary analog signal.

Suppose that an analog signal is comprised of polynomial segments of degree $n$: 
\begin{equation}
x\left(t\right) = \sum_{i}\sum_{j=0}^{n} x_{ij} \cdot \left(t-t_{i}\right)^j \cdot \left(u\left(t-t_{i}\right) - u\left(t-t_{i+1}\right)\right)
\end{equation}

Assuming that this signal is supplied as input to a system with an impulse response $f\left(t\right)$, the resulting output will be:
\begin{equation}
y\left(t\right) = \sum_{i}\sum_{j=0}^{n} x_{ij} \cdot \int_{t-t_{i+1}}^{t-t_i} f\left(\tau\right) \cdot \left(t-t_i-\tau\right)^j\,d\tau
\end{equation}

Applying the Binomial Theorem yields:
\begin{equation}
\begin{aligned}
y\left(t\right) &= \sum_{i}\sum_{j=0}^{n} x_{ij} \!\! \int\displaylimits_{t-t_{i+1}}^{t-t_i} \!\! f\left(\tau\right) \sum_{k=0}^{j}\binom{j}{k}\left(t-t_i\right)^{j-k}\left(-\tau\right)^k\,d\tau \\
&= \sum_{i}\sum_{k=0}^{n} c_{ik}\left(t\right)\cdot\left(F_k\left(t-t_i\right) - F_k\left(t-t_{i+1}\right) \right)
\end{aligned}
\end{equation}
where
\begin{equation}
\begin{aligned}
c_{ik}\left(t\right) &= \sum_{j=k}^{n} x_{ij} \cdot \binom{j}{k} \cdot \left(t-t_i\right)^{j-k} \\
F_k\left(t\right) &= \int_0^t f\left(\tau\right) \cdot \left(-\tau\right)^k\, d\tau
\end{aligned}
\end{equation}

The $n$ step response-like functions $F_k$ could be precomputed for use during emulation and, as before, these functions could be implemented using PWL tables. 

\subsection{Modeling Nonlinearity}

Memoryless nonlinearities occurring at the input or output of an analog block are straightforward to model in our architecture, since they can be implemented by lookup tables outside of the ADE.  For example, in our high-speed link implementation, the transfer function from the DCO code $n$ to the RX clock period $T_{\scriptscriptstyle RX}$ is given by the nonlinear relation $T_{\scriptscriptstyle  RX}=1/\left(\alpha+\beta n\right)$.  We used a small (160 bit) PWL lookup table to implement this behavior.

Our digitally-driven approach could also be used to model a block governed by the first-order nonlinear dynamics ${\dot{y} = g\left(x, y\right)}$, where $x$ is its piecewise-constant input and $y$ is its output.  Assuming that $x$ is constant from $t_1$ to $t_2$, the value of $y$ at the end of that time interval can be written as a function of the interval length and the values of $x$ and $y$ at the beginning of the interval:
\begin{equation}
y\left(t_2\right) = G\left(t_2-t_1, x\left(t_1\right), y\left(t_1\right)\right)
\end{equation}

In principle, the function $G$ could be precomputed for use during emulation.  While this approach might not always be practical, there are at least two cases that lend themselves to efficient implementation.

First, if the system's differential equation can be solved analytically for a constant input, then $G$ may have a convenient closed-form expression.  For example, an integrator that saturates to $\pm 1$ could be represented by:
\begin{equation}
y\left(t_2\right) = \mathrm{min}\left(\mathrm{max}\left(y\left(t_1\right) + \left(t_2-t_1\right)\cdot x\left(t_1\right), -1\right), 1\right)
\end{equation}

Second, if the input is restricted to certain discrete values, as in a high-speed link, then it may be possible to represent $G$ using a small number of precomputed trajectories.

As an example, consider a filter governed by the nonlinear dynamics ${\dot{y} = \left(x-y\right)/\tau\left(y\right)}$, with $\tau$ positive.  Assuming that the input is limited to values of $-1$ and $1$, there are effectively two unique output trajectories: one starting at $1$ and decreasing towards ${-1}$ (with input ${-1}$), and the other starting at ${-1}$ and increasing towards $1$ (with input $1$).  These two trajectories could be precomputed for use during emulation, so that whenever the input value changes, the trajectory corresponding to the new input value could be played back starting from the output value at the time of the input transition.

\section{Conclusion}

In this paper, we described an FPGA architecture for emulating a mixed-signal system with a digitally-driven analog block; that is, one whose input changes only on digital clock edges.  The analog output of such a block is computed as a weighted sum of pulse responses, and is calculated in a way that allows the emulator to progress directly from one digital clock edge to the next.  Unlike a conventional oversampled approach, the emulator's accuracy is independent of time step size. 

Using an \SI{8}{\giga\transaction/\second} high-speed link transceiver as an example, we implemented the proposed architecture on a Xilinx Zynq-7045 FPGA.  The emulation rate achieved was \SI{1.250}{\milli\second/\second}, which represents an 875x improvement over a high-performance CPU simulation of a similar system.  The worst-case error observed in comparison to an idealized CPU computation of the analog dynamics was -0.7/+1.1\%.

We conclude that the proposed architecture is appropriate for verifying the behavior of mixed-signal systems over long time scales, where CPU simulation would be impractically time-consuming.  Owing to its low resource utilization and ability to handle multiple clock domains, we expect that it will scale well to large designs.

\section*{Acknowledgment}
This work is supported by National Science Foundation Grant No. 1509126, a Hertz Foundation Fellowship, and a Stanford Graduate Fellowship.  

\bibliographystyle{ACM-Reference-Format}
\bibliography{iccad-bibliography}

\end{document}